\documentclass[aps,prl,twocolumn,superscriptaddress,showpacs,floats,preprintnumbers,amsmath,amssymb]{revtex4}

\usepackage{graphicx}
\usepackage{dcolumn}
\usepackage{bm}

\usepackage[pdffitwindow=false,bookmarks=true,pdfauthor={Weiss et al.},colorlinks=true,citecolor=blue,linkcolor=blue]{hyperref}
\usepackage{hypernat}
\usepackage{hypcap}

\begin{document}


\title{Imaging Pauli repulsion in scanning tunneling microscopy\\}

\author{C. Weiss}
\affiliation{Institut f{\"u}r Bio- und Nanosysteme 3,
Forschungszentrum J{\"u}lich, 52425 J{\"u}lich, Germany}
\affiliation{JARA--Fundamentals of Future Information Technology}
\author{C. Wagner}
\affiliation{Institut f{\"u}r Bio- und Nanosysteme 3,
Forschungszentrum J{\"u}lich, 52425 J{\"u}lich, Germany}
\affiliation{JARA--Fundamentals of Future Information Technology}
\author{C. Kleimann}
\affiliation{Institut f{\"u}r Bio- und Nanosysteme 3,
Forschungszentrum J{\"u}lich, 52425 J{\"u}lich, Germany}
\affiliation{JARA--Fundamentals of Future Information Technology}
\author{M. Rohlfing}
\affiliation{Fachbereich Physik, Universit{\"a}t Osnabr{\"u}ck,
49069 Osnabr{\"u}ck, Germany}
\author{F. S. Tautz}
\affiliation{Institut f{\"u}r Bio- und Nanosysteme 3,
Forschungszentrum J{\"u}lich, 52425 J{\"u}lich, Germany}
\affiliation{JARA--Fundamentals of Future Information Technology}
\author{R. Temirov}
\affiliation{Institut f{\"u}r Bio- und Nanosysteme 3,
Forschungszentrum J{\"u}lich, 52425 J{\"u}lich, Germany}
\affiliation{JARA--Fundamentals of Future Information Technology}

\date{\today}

\begin{abstract}
A scanning tunneling microscope (STM) has been equipped with a
nanoscale force sensor and signal transducer composed of a single
D$_{2}$ molecule that is confined in the STM junction. The
uncalibrated sensor is used to obtain ultra-high geometric image
resolution of a complex organic molecule adsorbed on a noble metal
surface. By means of conductance-distance spectroscopy and
corresponding density functional calculations the mechanism of the
sensor/transducer is identified. It probes the short-range Pauli
repulsion and converts this signal into variations of the junction
conductance.

\end{abstract}


\maketitle

\begin{figure}
\centering
\includegraphics[width=8cm]{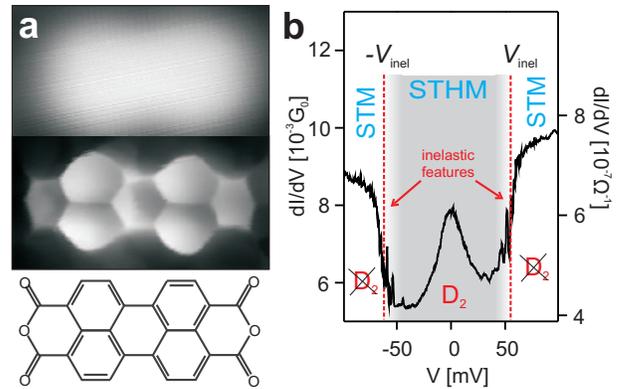}
\caption{(a) STM (top) and STHM image (bottom) of PTCDA
(3,4,9,10-perylenetetracarboxylic-dianhydride) on Au(111):
1.3$\times$0.7 nm$^{2}$, constant height, $D_{2}$, $V=316$~mV
(STM) and $V=-5$~mV (STHM). The chemical structure formula of
PTCDA is shown for comparison. (b) dI/dV spectrum measured in the
center of PTCDA/Au(111), recorded with lock-in detection (10~mV
modulation, frequency 2.3~kHz). $G_0=\frac{2e^2}{h}=
(12.9~k\Omega)^{-1}$ is the quantum of conductance.} \label{fig1}
\end{figure}

Since its invention the scanning tunneling microscope (STM) has
become an important tool of nanoscience, because it routinely
provides {\AA}ngstr\"om-scale image resolution on various sample
surfaces \cite{1,2,3,4,5,corieder}. However, STM suffers from a
serious drawback---the inability to resolve complex chemical structure. 
This disadvantage arises because the STM probes the local density of
states (LDOS) in the vicinity of the Fermi level, while details of
the chemical structure are primarily encoded in lower-lying
orbitals. Better access to chemical structure is therefore
provided by mapping the total electron density (TED). Indeed, it
has been shown recently that non-contact atomic force microscopy
is able to resolve the inner structure of a complex organic
molecule, by imaging short-range repulsive interactions that
depend on the TED \cite{6}. Even earlier, however, it has been
demonstrated that STM acquires very similar force imaging
capabilities when operated in the so-called scanning tunneling
hydrogen microscopy (STHM) mode \cite{7}---STHM images indeed
closely resemble the chemical structure formulae of the
investigated compounds (see ref.~\cite{7} and Fig.\ref{fig1}a). In
this letter we present an analysis of the STHM junction by means
of $dI/dV$(z)-spectroscopy and density functional theory that
allows us to explain its imaging mechanism for the first time.

The experiments were performed on PTCDA/Au(111) with a Createc
low-temperature STM operated at 5-10~K in ultra-high vacuum (UHV).
The electrochemically etched W tips and the Au(111) surface have
been prepared using Ar$^{+}$ sputtering and temperature annealing
in UHV. The STM tips were additionally prepared by indentation
into the clean gold surface. PTCDA molecules were deposited from a
quartz crucible mounted in a home-built Knudsen cell. Deposition
of H$_{2}$ or D$_{2}$ was performed according to the recipe
described in ref.~\cite{7}. Since both H$_{2}$ and D$_{2}$ yield
similar results, we restrict the discussion in this letter to
D$_{2}$.

We start by summarizing the features of STHM that have been
reported before \cite{7}. The best STHM resolution (as shown in
Fig.~\ref{fig1}a) is achieved in constant-height mode with rather 
low tunneling bias
$|V|\lesssim 10$~mV. The appearance of STHM imaging after D$_{2}$
dosing coincides with the emergence of non-linear differential
conductance spectra $G(V)\equiv dI/dV(V)$ close to zero bias. In
particular, inelastic features appear (in Fig.~\ref{fig1}b at $\pm
V_\mathrm{inel}$) \cite{7,8,9,10}. In some cases, as in
Fig.~\ref{fig1}b, a pronounced zero bias anomaly has also been
observed \cite{7}. Imaging in the presence of D$_{2}$ can be
switched reversibly between the STM and STHM modes by changing the
applied bias $V$ \cite{7}: If $|eV|$ exceeds $|eV_\mathrm{inel}|$,
the junction images in the conventional STM (or LDOS) mode. If,
however, $|eV|$ is sufficiently smaller than $|eV_\mathrm{inel}|$,
the junction operates in the STHM mode, yielding images with
ultra-high geometric resolution. The nature of the inelastic
features has been investigated in detail in nanojunctions
containing H$_{2}$ \cite{7,8,9,10,11}. They are assigned to
transitions between two structurally different states of the
junction. However, the precise nature of the associated two-level
system is still debated \cite{8,9,10}.

Understanding the structure of the STHM junction is vital for
identifying the imaging mechanism. Based on the conductance values
in our experiments ($>5\times 10^{-5}\mathrm{G_{0}}$) we conclude
that the tip-surface separation must be smaller than 1~nm
\cite{22}. At such distances the junction can accommodate only a
single monolayer of D$_{2}$ \cite{20,14}. Clearly, the
deuterium-induced imaging mode requires the presence of D$_{2}$ just
below the tunneling tip apex. Given
the high resolution that is achieved in STHM, which in
Fig.~\ref{fig1}a is of the order 50~pm, at most one D$_{2}$
molecule can be in the active area of the junction. Therefore
we can model the junction in the STHM mode, i.e. at
$|V|<|V_\mathrm{inel}|$, by a single D$_{2}$ molecule physisorbed
\cite{19} between the tip apex and the sample surface. For
$|V|>|V_\mathrm{inel}|$, the D$_{2}$ molecule is displaced away
from the tip apex, as proven by the recurrence of conventional STM
imaging in conjunction with the occurrence of a structural
transition at $|V_\mathrm{inel}|$.

\begin{figure}
\centering
\includegraphics[width=8 cm]{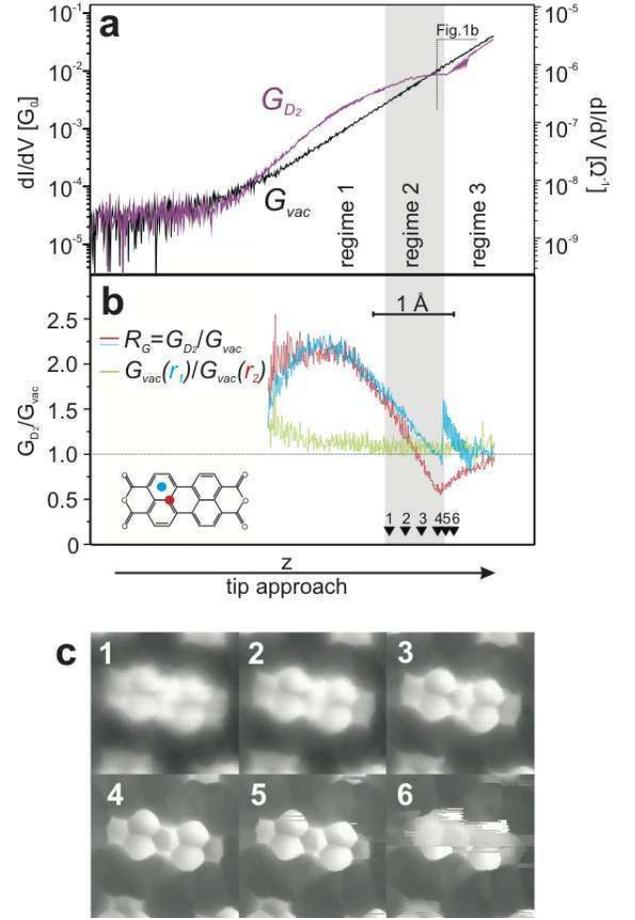}
\caption{(a) $\frac{dI}{dV}(z)$ spectra recorded at fixed bias,
measured with $D_{2}$ on PTCDA/Au(111) ($z$-axis scale given by
scale bar). $G_0=\frac{2e^2}{h}$ is the quantum of conductance.
$G_{D_2}$ (magenta): $V=-5$~mV. $G_{vac}$
(black): $V=-130$~mV. All spectra recorded from the same
stabilization point. (b) $R_\mathrm{G}(z)$ curves (cf. text) measured at
positions $\vec{r}_1$ (blue) and $\vec{r}_2$ (red) marked in the
lower inset. The ratio
$G_{vac}(\vec{r}_1)/G_{vac}(\vec{r}_2)$ is shown
in green. (c) STHM images ($1.3 \times 1.3 $ nm$^{2}$, constant height, $V=-5$~mV) of
PTCDA/Au(111) measured with D$_{2}$ at different $z$ indicated by
black triangles in panel b.} \label{fig2}
\end{figure}

Switching the bias between $|V|<|V_\mathrm{inel}|$ and
$|V|>|V_\mathrm{inel}|$ we can study the properties of a given
junction, and in particular the differential conductance $G$ as a
function of tip-surface distance $z$, in presence and absence of
the confined D$_{2}$ molecule, without any other structural
changes of the junction. Fig.~\ref{fig2}a displays the results of
$G(z)$-spectroscopy for the empty junction (curve measured with
$|V|>|V_\mathrm{inel}|$ and labelled $G_{vac}$) and for
the junction with D$_{2}$ (curve measured with
$|V|<|V_\mathrm{inel}|$ and labelled $G_{D_2}$).
$G_{vac}$ increases exponentially with decreasing $z$,
as expected for vacuum tunneling. In contrast, $G_{D_2}$
behaves non-exponentially. Two opposing tendencies are observed:
For intermediate $z$, $G_{D_2}$ exceeds
$G_{vac}$, while for smaller tip-surface distances the
situation reverses. Based on the data in Fig.~\ref{fig2}a, we can
define three characteristic regimes.

At the shortest distances recorded in Fig.~\ref{fig2}a (regime 3)
$G_{D_2}$ exhibits increased noise and conductance
jumps. Such conductance changes usually occur when the tunneling
junction undergoes structural modifications due to the tip
contacting the sample surface. Notably, the discontinuities in
$G_{D_2}$ occur at tip-surface distances where
$G_{vac}$ still behaves strictly exponentially, i.e.~the
empty junction is still out of contact. Hence, the contact in
question must occur via the confined D$_{2}$ molecule. Clearly,
the associated structural changes will occur in the softest part
of the junction, i.e.~the D$_{2}$ molecule, which eventually is
squeezed out of the junction. Indeed, images measured at the onset
of regime 3 (images 5 and 6 in Fig.~\ref{fig2}c) show a sudden loss of
STHM resolution.

Having identified regime 3 with the squeezing of the D$_{2}$
molecule out of the junction, we can associate the preceding
regime 2 with the gradual compression of the junction that
eventually causes this squeeze-out. To quantify the effect of 
D$_{2}$ on the junction conductance in
regimes 1 and 2, we define the \emph{conductance ratio}
$R_\mathrm{G}(z)\equiv G_\mathrm{D_2}(z)/G_\mathrm{vac}(z)$.
As can be seen in Fig.~\ref{fig2}a, $R_\mathrm{G}(z)$ can be
larger or smaller than 1, depending on the value of $z$. 
Fig.~\ref{fig2}b (red, blue) shows that in
regime 2, where the best STHM resolution is recorded (images 3 to
5 in Fig.~\ref{fig2}c), $R_\mathrm{G}(z)$ curves measured at
different lateral positions above the PTCDA molecule each display
a distinct slope. At the same time $G_{vac}$ curves 
do not vary appreciably from point to point above PTCDA
(Fig.~\ref{fig2}b, green), which is consistent with the blurred
and featureless STM images that are recorded with the empty
junction at $|V|>|V_\mathrm{inel}|$ (Fig.~\ref{fig1}a). The
$R_\mathrm{G}(z)$ curves in Fig.~\ref{fig2}b hence show that
\emph{the STHM contrast in Fig.~\ref{fig1}a can be ascribed 
to the effect of D$_2$ on the junction conductance which becomes 
pronounced in a narrow range of $z$-values in regime 2}. We therefore
need to study the nature of regime 2 in more detail.

\begin{figure}
\centering
\includegraphics[width=7 cm]{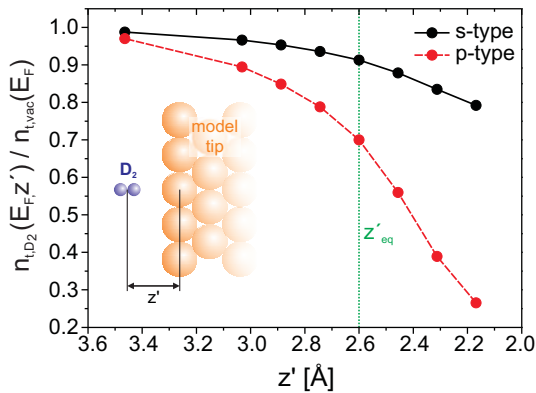}
\caption{ DFT-LDA simulated
$n_\mathrm{t,D_2}(E_F,z')/n_\mathrm{t,vac}(E_F)$ vs. D$_2$-tip
distance $z'$ for $s$- (black) and $p$-type (red) orbitals at the
Au atom below the deuterium molecule. $n_\mathrm{t,D_2}(E_F,z')$ is the
LDOS of the model tip at the Fermi level at given D$_2$-tip
distance $z'$ and $n_\mathrm{t,vac}(E_F)$ is LDOS at the
Fermi level of the bare tip electrode. Equilibrium distance
$z'_\mathrm{eq}$ is indicated by the vertical line. For details of
the simulation, cf.~ref.~\cite{calculationdetails}. }\label{fig3}
\end{figure}

To investigate the influence of D$_2$ on the tunnelling conductance of an
STHM junction in regime 2, we have carried out a density functional theory (DFT)
calculation in which we systematically varied the distance $z'$ 
between D$_2$ and an Au(111) surface (Fig.~\ref{fig3} inset) \cite{calculationdetails}. 
This surface was chosen to model the tip electrode in our STHM experiments.
We calculate the model tip DOS at the Fermi level, $n_\mathrm{t,D_2}(E_F,z')$,  
as the LDOS at the metal atom located directly underneath the D$_2$. 
Dividing  $n_\mathrm{t,D_2}(E_F,z')$ by $n_\mathrm{t,vac}(E_F)$, 
the LDOS of the bare tip without D$_2$,  
we find that the tip DOS decays substantially with
decreasing $z'$ (Fig.~\ref{fig3}). The origin of this behavior is
the Pauli exclusion principle: To minimize overlap between the
closed shell of D$_2$ and the metal electrons, both of their wave
functions must rearrange locally, which depletes the metal's local
DOS in the vicinity of the Fermi level \cite{12,13,14}, while the
associated energy cost leads to a repulsive force between D$_2$
and the metal (Pauli repulsion). 

In the limit of low tunneling bias Tersoff-Hamann theory of STM 
predicts $G\propto n_\mathrm{t,D_2}(E_F,z')n_\mathrm{s}(E_F,\vec{r}_\mathrm{t})$, where 
$n_\mathrm{s}$ is the sample LDOS at tip position $\vec{r}_\mathrm{t}$ \cite{17}.
Accordingly, $G$ and $R_\mathrm{G}$ must decrease proportionally to $n_\mathrm{t,D_2}(E_F,z')$ 
as D$_2$ approaches the tip. DFT results shown in Fig.~\ref{fig3} suggest that
the rate of tip DOS decrease should be in the range from 0.2 to 1 $\mathrm{\AA}^{-1}$.
At the same time experimental data from Fig.~\ref{fig2}b show rates 
between 1.1 and 1.5 ~\AA$^{-1}$. To be able to compare both results we
additionally have to  divide experimental values by the factor $\frac{dz'}{dz}$ 
accounting for the differences between the scales $z$ and $z'$. $\frac{dz'}{dz}$ 
must be in the range between $\simeq 0$ (soft sample) to 1 (hard wall sample).
Given the effects of the unknown tip shape and the neglect of modifications
of the sample LDOS by Pauli repulsion \cite{yy}, the agreement
between experiment and simulation is remarkable and we can thus
conclude that \emph{the $z$-variation of $R_\mathrm{G}(z)$ in
regime 2 can be explained as an effect of Pauli repulsion in the STHM
junction}.

\begin{figure}
\centering
\includegraphics[width=5.4cm]{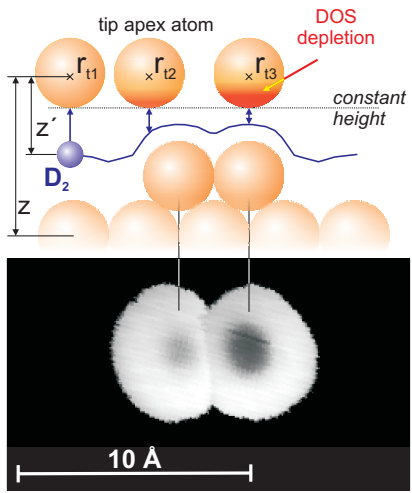}
\caption{(a)(b) STHM image of a Au dimer on Au(111) (constant
height, $V=$2~mV,  $D_{2}$ ) (bottom panel) and schematic sketch
of contrast generation. Cf.~text for details.} \label{fig4}
\end{figure}

This result holds the key for understanding the contrast formation
in STHM. To demonstrate this, we first discuss the STHM contrast
above an inherently simple object, namely Au adatoms on Au(111),
before turning to the more complex PTCDA molecule with its
internal structure. Fig.~\ref{fig4} (bottom panel) shows the
experimental image of an Au adatom dimer. The image has been
recorded at constant height, nearly zero bias voltage (2~mV), and
with a junction containing D$_2$, i.e.~under STHM conditions. We
observe two well-separated structures, each of which corresponds
to one of the adatoms. In comparison to the flat sample surface,
the adatoms appear bright, i.e.~with a large $G(\vec{r}_\mathrm{t})$, because the sample LDOS
$n_\mathrm{s}(\simeq E_F,\vec{r}_\mathrm{t})$ at tip positions close
to the adatoms, e.g.~at $\vec{r}_\mathrm{t2}$, is increased with
respect to the one at $\vec{r}_\mathrm{t1}$, due to a reduced
effective tip-sample distance (cf.~the top panel of
Fig.~\ref{fig4}). So far this is not different from conventional
constant height imaging in STM. However, in the center of each of
the adatoms (i.e.~at $\vec{r}_\mathrm{t}\simeq
\vec{r}_\mathrm{t3}$) a dark area is observed in Fig.~\ref{fig4}. 
In contradiction to conventional STM \cite{bbb}, Fig.~\ref{fig4} clearly implies 
that $G(\vec{r}_\mathrm{t3})< G(\vec{r}_\mathrm{t2})$.
The reason for this deviation from the normal STM behavior can be
found in the presence of D$_2$ in the junction, and in particular
in its trajectory, which is displayed schematically in Fig.~\ref{fig4} (top): As the
tip moves from $\vec{r}_\mathrm{t2}$ to $\vec{r}_\mathrm{t3}$, the
D$_{2}$ molecule will have to move to a new vertical
equilibrium position closer to the tip (smaller $z'$) because at that position the stronger Pauli
repulsion from the adatom is balanced by a stronger Pauli repulsion
from the tip. In conjunction with Fig.~\ref{fig3}, this must lead 
to a sharp reduction in $n_\mathrm{t}(\simeq E_F)$. 
According to Fig.~\ref{fig4} this reduction overcompensates the
rise in $n_\mathrm{s}(\simeq E_F,\vec{r}_\mathrm{t})$
\cite{yy} and leads to the dark areas in the centers of the
adatoms. The analysis of the adatom image thus
shows that \emph{the STHM contrast can be understood as
an $(x,y)$-map of the short-range Pauli repulsion from the sample
surface acting on the D$_{2}$ molecule in the STHM junction, superimposed 
over the conventional LDOS contrast}.

With this knowledge, we can finally analyze the STHM contrast
generation above PTCDA. As in the case of the dimer, the D$_{2}$
molecule follows the tip and probes lateral variations of the
Pauli repulsion from the adsorbed PTCDA. For example, when the tip
moves from a position above the center of a C$_{6}$ ring to a
position directly above a carbon atom, the D$_{2}$ molecule in the
junction will---similarly to the trajectory shown in
Fig.~\ref{fig4}---move to a higher equilibrium position closer to
the tip, because of the increased TED above the carbon atom. As in
the case of the adatoms discussed above, this leads to a reduced
$n_\mathrm{t}(\simeq E_F)$ and therefore lower conductance. In the STHM image
the carbon atoms of PTCDA (and by a similar argument the
$\sigma$-bonds between the carbons) therefore appear darker than
the "empty" spaces inside the C$_6$ and C$_5$O rings of the PTCDA
backbone, just as observed in the image of Fig.~\ref{fig1}a.

In conclusion, we arrive at the following model of STHM imaging: A
single D$_{2}$ molecule is physisorbed in the STM junction, such
that it is confined directly underneath the tip apex. This
molecule is the crucial element in the STHM imaging process, as it
probes the short-range Pauli repulsion from the surface (sensor
action) and transforms this force signal into variations of the
tunneling conductance (transducer action), the latter again via
Pauli repulsion. Because of its nanoscale size, the sensor is
insensitive to long-range forces. Clearly, the sensor/transducer
modulates the tunneling conductance on top of the normal LDOS
contrast. As long as the Pauli-induced conductance modulation 
is larger than the LDOS-induced change of the
background conductance itself, the image will be dominated by the
STHM contrast. We note that the
described functionality should also work with other closed-shell
particles besides hydrogen and deuterium \cite{EiglerXe}.
Comparing STHM to conventional STM, direct tunneling between tip
and sample surface still forms the basis of imaging in STHM.
However, in STHM a compliant element that is sensitive to a
laterally varying sample property other than the LDOS is added to
the tunneling junction, where it modulates the tunneling current
that is used for imaging.

Financial support from the Helmholtz Gemeinschaft is gratefully
acknowledged, as are helpful discussions with J. Kroha (Bonn), 
S. Bl\"ugel and N. Atodiresei (J\"ulich).


\begin{thebibliography}{99}
\bibitem{22}
At contact, which corresponds to atom-atom distances of
approximately 3~\AA, typical conductances are of the order
G$_{0}$. For a typical metal,  the tunneling current decays by a
factor of 7.4 per \AA ngstr\"om \cite{1}. For our minimum
conductance of $5\times 10^{-5}\mathrm{G_{0}}$ this yields a
maximum distance of 8~\AA.
\bibitem{calculationdetails}
Geometry optimization for D$_2$ on Au(111) was done in a $3\times
3$ overlayer on a 3-layer slab within the local density
approximation, yielding an upright D$_2$ molecule in an on-top
position with an equilibrium distance
$z'_\mathrm{eq}=2.6~\mathrm{\AA}$. For the LDOS calculation, D$_2$
molecules in this adsorption geometry, but in a $\sqrt{3}\times
\sqrt{3}$ overlayer with an 11-layer slab for better convergence,
were moved in steps of $0.15~\mathrm{\AA}$ towards and away from
the surface. For each configuration, the LDOS was calculated by
projecting the quantum mechanical states on the local orbitals
($s$, $p_z$, $d_{z^2}$) of the Au atom directly below the D$_2$.
The $d_{z^2}$-orbital does not yield a significant contribution.
Note that DFT does not consider the masses of the nuclei (unless
dynamics of the nuclei is considered), i.e.~in DFT there is no
difference between H$_2$ and D$_2$.
\bibitem{yy}
For simplicity, we have neglected here the influence of D$_2$ on
$n_\mathrm{s}$. This would further amplify the variation of
$G(\vec{r})$.
\end{thebibliography}
\end{document}